\documentclass[twocolumn,pre,superscriptaddress]{revtex4-1}
\usepackage{float}
\usepackage{graphicx}
\usepackage{amssymb,amsmath}
\usepackage{bm}
\usepackage{dcolumn}

\usepackage{hyperref}
\hypersetup{backref,colorlinks=true,breaklinks,urlcolor=blue,linkcolor=blue,citecolor=blue}
\usepackage{color}

\usepackage[capitalise]{cleveref}
\usepackage[notrig]{physics}   
\usepackage{braket}

\newcommand{\rom}[1]{\uppercase\expandafter{\romannumeral #1}}

\begin{document}

\title{Lindbladian approximation beyond ultra-weak coupling}
\author{Tobias Becker}
\email{tobias.becker@tu-berlin.de}
\affiliation{Institut f\"ur Theoretische Physik, Technische Universit\"at Berlin, Hardenbergstrasse 36, 10623 Berlin, Germany}
\author{Ling-Na Wu}
\email{lingna.wu@tu-berlin.de}
\affiliation{Institut f\"ur Theoretische Physik, Technische Universit\"at Berlin, Hardenbergstrasse 36, 10623 Berlin, Germany}
\author{Andr\'{e} Eckardt}
\email{eckardt@tu-berlin.de}
\affiliation{Institut f\"ur Theoretische Physik, Technische Universit\"at Berlin, Hardenbergstrasse 36, 10623 Berlin, Germany}

\begin{abstract}
Away from equilibrium, the properties of open quantum systems depend on the details of their environment. A microscopic derivation of a master equation (ME) is therefore crucial. Of particular interest are Lindblad-type equations, not only because they provide the most general class of Markovian MEs, but also since they are the starting point for efficient quantum trajectory si\-mu\-la\-tions. Lindblad-type MEs are commonly derived from the Born-Markov-Redfield equation via a rotating-wave approximation (RWA). However the RWA is valid only for ultra-weak system bath coupling and often fails to accurately describe nonequilibrium processes. Here we derive an alternative Lindbladian approximation to the Redfield equation, which does not rely on ultra-weak system-bath coupling. Applying it to an extended Hubbard model coupled to Ohmic baths, we show that, especially away from equilibrium, it provides a good approximation in large parameter regimes where the RWA fails.
\end{abstract}

\maketitle
\section{Introduction} Quantum systems are inevitably interacting with their surrounding environment and very often this effect needs to be taken into account for an accurate description of their properties. In equilibrium the weak coupling to a thermal environment can be described by statistical mechanics, i.e.\ Gibbs ensembles, without considering the details of the environment beyond a few thermodynamic variables like temperature and chemical potential. However, very often we are interested in quantum systems far from thermal equilibrium, for instance, in the context of quantum information processing \cite{SAlipour2020,SVinjanampathy2020}, when considering quantum heat engines \cite{MEsposito2010,MParanauLlobet2020}, or when controlling quantum matter via strong driving \cite{TShirai2016,MSato2020}. In general this is a non-trivial regime in which the properties of a quantum system depend on the details of the environment. Since, a full description of a large environment is typically neither of interest nor feasible, the system is usually described within the theory of open quantum systems \cite{breuerpetruccione} by (microscopically) deriving a master equation (ME). In the Markovian case, where memory effects are negligible, the Lindblad ME is the most general quantum ME \cite{GLindblad1976,GoriniKossakowski1976}. It is also the the basis for the efficient stochastic simulation of larger systems by means of quantum trajectories \cite{Dalibard92,ADaley2014,KMolmer1993,GCHegerfeldt1991,GLueders1951}. \newline 
The standard approach for microscopically deriving a ME is the Born-Markov approximation leading to the Redfield ME \cite{FBloch56,AGRedfield65}. See for example the recent developments in quantum chemistry \cite{JThingaPHaenggi2012,EFranciscoMLRonald92,AMCastilloDRReichman15}, atomic physics \cite{FDamanetAJDaley19} and quantum optics \cite{PHaenggi05,Strunz20,Alicki18}. 
From this a Lindblad ME follows, when further employing the rotating-wave approximation (RWA) \cite{PHaenggi2005,breuerpetruccione,carmicheal,CWGardiner00}. However, this additional step requires ultra-weak coupling (which is small compared to the energy level splitting of the system) and the RWA only predicts the correct steady-state in the zeroth order of the coupling \cite{JThingaPHaenggi2012}. Problems of the RWA also become significant in the transient dynamics \cite{BBalzerGStock2004,XXuJThinga2019,ADaley2019} as well as for transport properties \cite{HWichterich2007}. \newline
In recent years there has been ongoing development for Lindblad approximations that bypass the RWA \cite{GSchallerTBrandes2008,PhysRevLett.104.070406,HallCresserAndersson2014,PhysRevB.97.035432,Mozgunov2020completelypositive,PhysRevB.101.125131,NathanRudner2020}. 
In this work we provide a general approach for deriving an alternative Lindbladian approximation to the Redfield equation that is valid also in regimes of finite coupling, where the RWA fails. It is based on an optimized diagonalization of the Redfield dissipator. We test the resulting ME for an extended Hubbard chain coupled to Ohmic baths and show that in a large pa\-ra\-me\-ter regime where the RWA fails, it provides an accurate description of the Redfield dynamics. Combining our approach with quantum trajectory simulations, we are able to simulate system sizes that we cannot treat by integrating the Redfield equation.
\section{Redfield equation} The starting point for our approach is the Redfield formalism. The total Hamiltonian for the system-bath compound reads  $\hat{H}_\mathrm{tot}=\hat{H}_\mathrm{S} + \hat{H}_\mathrm{SB} + \hat{H}_\mathrm{B}$, with the system and bath Hamiltonian $\hat{H}_\mathrm{S}$ and $\hat{H}_\mathrm{B}$, respectively. The interaction between system and bath is described by $\hat{H}_\mathrm{SB} = \hat{S}\otimes\hat{B}$, where $\hat{B}$ shall carry the dimension of energy and $\hat{S}$ is a dimensionless hermitian operator acting on the system. The case of several independent baths and non-hermitian coupling is  outlined in \cref{sec:multiplebaths}.  \newline
The time evolution of the reduced density matrix for the system $\hat{\rho}=\tr_\mathrm{B}(\rho_\mathrm{tot})$ shall be described in Born-Markov approximation \cite{breuerpetruccione,TAlbash2012}. First, the Born approximation provides a factorization of system and bath states, i.e.\ $\hat{\rho}_\mathrm{tot} = \hat{\rho}\otimes\hat{\rho}_\mathrm{B}$, where the bath stays in thermal equilibrium $\hat{\rho}_\mathrm{B}=\exp[-\beta\hat{H}_\mathrm{B}]/\tr_\mathrm{B}\exp[-\beta\hat{H}_\mathrm{B}]$ at an inverse temperature $\beta=1/T$. Additionally, in the Markov approximation bath correlations are assumed to decay fast compared to the time scales of the system dynamics, resulting in the time-local time-dependent Redfield ME \cite{FBloch56,AGRedfield65},
\begin{align}
\begin{split}
	\dot{\hat{\rho}}  &= -\frac{i}{\hbar} [ \hat{H}_\mathrm{S} , \hat{\rho} ] + \hat{S}\hat{\rho}\hat{\mathbb{S}}_t^\dagger + \hat{\mathbb{S}}_t \hat{\rho}\hat{S}  -  \hat{S} \hat{\mathbb{S}}_t \hat{\rho}-  \hat{\rho}\hat{\mathbb{S}}_t^\dagger\hat{S} , \\
	\hat{\mathbb{S}}_t &= \int\limits_0^{t} C_\tau\, \hat{S}_{-\tau}\ d\tau,
\end{split}
\label{eq:redfield}
\end{align} 
with bath correlation $C_\tau=\tr_\mathrm{B}(\hat{B}_\tau\hat{B}\hat{\rho}_\mathrm{B})/\hbar^2$ and Heisenberg operators ${\hat{S}_{\tau}=\exp[i\hat{H}_\mathrm{S} \tau/\hbar] \hat{S}\exp[-i \hat{H}_\mathrm{S}\tau/\hbar]}$ and $\hat{B}_{\tau}=\exp[i\hat{H}_\mathrm{B} \tau/\hbar] \hat{B}\exp[-i \hat{H}_\mathrm{B}\tau/\hbar]$. Often a further approximation is made by setting $\hat{\mathbb{S}}_t \approx \hat{\mathbb{S}}_\infty$, which is sufficient for the late-time or steady-state behaviour \cite{XXuJThinga2019}. In contrast, we will keep the time dependence. The last two terms of the Redfield equation (\ref{eq:redfield}) are not purely dissipative but also contribute to the coherent dynamics. We split the Redfield equation (\ref{eq:redfield}) into coherent and dissipative part as $\dot{\hat{\rho}} = (-i/\hbar) [\hat{H}_\mathrm{S} + \hat{H}_t^\mathrm{LS}, \hat{\rho}] + \mathcal{D}_t^\mathrm{Red}[\hat{\rho}]$, with Lamb-shift Hamiltonian and Redfield dissipator
\begin{align}
\hat{H}_t^\mathrm{LS}& =\hbar \frac{\hat{S} \hat{\mathbb{S}}_t-\hat{\mathbb{S}}_t^\dagger \hat{S}}{2i} \label{eq:lambshift},\\
\mathcal{D}_t^\mathrm{Red}[\hat{\rho}] &=\hat{S} \hat{\rho} \hat{\mathbb{S}}_t^\dagger + \hat{\mathbb{S}}_t \hat{\rho}\hat{S}  -  \frac{1}{2}\Big\{ \hat{S} \hat{\mathbb{S}}_t + \hat{\mathbb{S}}_t^\dagger \hat{S}, \hat{\rho} \Big\}
\label{eq:redfielddissipator}
\end{align}
respectively, where $\{.,.\}$ denotes the anti-commutator. The Redfield dissipator (\ref{eq:redfielddissipator}) is not of Lindblad-form~\cite{breuerpetruccione,GoriniKossakowski1976,HWichterich2007} as will be seen also explicitly from equation (\ref{eq:pseudolindblad}) below. 
\section{Rotating-wave approximation} The standard way to derive a Lindblad ME is closely related to the representation of the Redfield dissipator in the eigenbasis of $\hat{H}_\mathrm{S}$, $S_{qk}=\matrixelement{q}{\hat{S}}{k}$ and $\hat{H}_\mathrm{S}\ket{q}=\varepsilon_q \ket{q}$. For later convenience let the coupling matrix fulfill the normalization condition $\sum_{qk} |S_{qk}|^2=1$, i.e.\ the coupling strength is absorbed in the bath operator $\hat{B}$. In the eigenbasis the Heisenberg operator takes the form $\hat{S}_{-\tau}= \sum_{qk} S_{qk} \exp[-i\Delta_{qk}\tau/\hbar] \hat{L}_{qk}$, with jump operators $\hat{L}_{qk}=\ketbra{q}{k}$ and level splitting $\Delta_{qk}=\varepsilon_q - \varepsilon_k$. The Redfield equation (\ref{eq:redfield}) is quadratic in $\hat{S}$ and, thus, runs over four indices $q,k$ and $q',k'$ for the two pairs of level splittings $\Delta_{qk}$ and $\Delta_{q'k'}$. For very weak coupling the oscillations of the non-secular terms with $\Delta_{qk}\ne\Delta_{q'k'}$ are much faster than the slow variation of the state induced by the coupling and, thus, average out. Neglecting all but the terms with $\Delta_{qk}=\Delta_{q'k'}$ leads to the RWA \cite{breuerpetruccione,carmicheal,CWGardiner00},
\begin{align}
\hat{H}_t^\mathrm{LS,RWA}&= \sum_{qk} \hbar\,h_t(\Delta_{qk})|S_{qk}|^2 \hat{L}_{qk}^\dagger\hat{L}_{qk},\\
\mathcal{D}_t^\mathrm{RWA}[\hat{\rho}] &= \sum\limits_{qk} 2\, g_t(\Delta_{qk})|S_{qk}|^2 \Big[ \hat{L}_{qk} \hat{\rho} \hat{L}_{qk}^\dagger-  \frac{1}{2} \Big\{ \hat{L}_{qk}^\dagger \hat{L}_{qk} , \hat{\rho}  \Big\} \Big], 
\label{eq:RWA}
\end{align}
where $g_t$ and $h_t$ denote the real and imaginary part of the bath correlation function $G_t(\Delta)=\int_0^t \exp[-i\Delta\tau/\hbar]\, C_\tau\, d\tau$. The Lamb-shift Hamiltonian $\hat{H}_t^\mathrm{LS,RWA}$ is diagonal in the energy-basis ($\hat{L}_{qk}^\dagger\hat{L}_{qk}=\ketbra{k}{k}$) and thus modifies the coherent dynamics only by shifting the eigenenergies. In turn, the dissipator is of Lindblad-form and describes quantum jumps between individual energy eigenstates. One also obtains decoupled equations of motion for the diagonal and off-diagonal entries of the density matrix. The off-diagonals decay exponentially leading to a diagonal steady-state. For a thermal bath at inverse temperature $\beta$ this is of canonical Gibbs form, i.e.\ $\hat{\rho}_{ss}^\mathrm{RWA}=\exp[-\beta\hat{H}_\mathrm{S}]/\tr_\mathrm{S}\exp[-\beta\hat{H}_\mathrm{S}]$. This is independent of the coupling $\hat{H}_\mathrm{SB}$ and therefore it only captures the zero coupling limit \cite{JThingaPHaenggi2012,PHaenggi2005}. 
\section{Optimized truncation approach} For weak but finite coupling, where the RWA fails, we now derive an alternative approximation to the Redfield equation, which also leads to a Lindblad ME. For this purpose, we first bring the Redfield dissipator (\ref{eq:redfielddissipator}) into the diagonal form
\begin{align}
\mathcal{D}_t^\mathrm{Red}[\hat{\rho}]  =  \sum_{\sigma=+,-} \sigma \Big[\hat{A}_t^\sigma \hat{\rho} \hat{A}_t^{\sigma \dagger} - \frac{1}{2} \Big\{ \hat{A}_t^{\sigma\dagger} \hat{A}_t^\sigma , \hat{\rho}  \Big\} \Big],
\label{eq:pseudolindblad}
\end{align}
by introducing the new jump operators
\begin{equation}
\hat{A}_t^{\pm} = \frac{1}{\sqrt{2 \cos \varphi_t }}  \Big[\lambda_t^\pm \hat{S}\pm \frac{1}{\lambda_t^\pm} \hat{\mathbb{S}}_t\Big],
\label{eq:jumpoperators}
\end{equation}
with $\lambda_t^\pm=\lambda_t \exp{(\mp i \frac{\varphi_t}{2})}$  and arbitrary real, time dependent parameters $\lambda_t$ and $\varphi_t$, and where $\lambda_t^{-2}$ carries the dimension of time. By plugging \cref{eq:jumpoperators} into \cref{eq:pseudolindblad} in \cref{sec:pseudoLindblad} it is verified that these equations provide an exact representation of the Redfield dissipator~(\ref{eq:redfielddissipator}). The freedom of choosing $\lambda_t$ and $\varphi_t$ will be crucial in the following. Since the prefactor of the second term in \cref{eq:pseudolindblad} is negative, we refer to it as pseudo-Lindblad dissipator. A similar decomposition of the Redfield equation has been used recently in reference \cite{CGneiting2020} however for a specific choice of $\lambda_t^\pm$ which does not correspond to the optimal value that we derive below. Dissipators of the type of \cref{eq:pseudolindblad} are also used for time-convolutionless description of non-Markovian processes \cite{SAlipour2020,breuer2004,Piilo2009}. In contrast to the RWA, \cref{eq:pseudolindblad,eq:jumpoperators} are obtained without diagonalizing the system's Hamiltonian. Also, whereas in the RWA the number of jump operators grows quadratically with the Hilbert space dimension the pseudo-Lindblad dissipator \cref{eq:pseudolindblad} only has two jump operators. \newline
Finally \cref{eq:pseudolindblad} is reduced to Lindblad-form by neglecting the negative contribution,
\begin{align}
\mathcal{D}_t^\mathrm{Red}[\hat{\rho}] \simeq\mathcal{D}_t^\mathrm{trunc}[\hat{\rho}]  = \hat{A}_t^+ \hat{\rho} \hat{A}_t^{+ \dagger} - \frac{1}{2} \Big\{ \hat{A}_t^{+\dagger} \hat{A}_t^+ , \hat{\rho}  \Big\}.
\label{eq:truncatedME}
\end{align}
This truncation can be expected to be justified as long as the weight of the negative contribution $\lVert \hat{A}_t^- \rVert^2$ is small compared to the weight of the positive contribution $\lVert \hat{A}_t^+ \rVert^2$. In the following, we will compute the weight using the Frobenius norm $\lVert \hat{A}_t^\pm \rVert^2 = \mathrm{tr}_\mathrm{S}(\hat{A}_t^\pm \hat{A}_t^{\pm\dagger})$. \newline
Due to the special form of the jump operators $\hat{A}_t^\pm$, the optimal values for $\lambda_t$ and $\varphi_t$ minimize the weight of the negative contribution both absolutely and relative to the positive contribution. The optimization is carried out in \cref{sec:optimization} and one finds the optimal values $\lambda_t^4=\overline{g_t^2} + \overline{h_t^2}$ and $\sin\varphi_t=\overline{h_t}/(\overline{g_t^2} + \overline{h_t^2})^{1/2}$, where the overline denotes an average defined by $\overline{x} =\sum_{qk} x(\Delta_{qk}) |S_{qk}|^2$. Here $|S_{qk}|^2$, with $\sum_{qk} |S_{qk}|^2=1$, plays the role of a probability distribution. We could interpret these results, e.g.\ by identifying $\lambda_t^{-2}$ with the typical timescale that is related to the amplitude of the bath correlation function. The optimization is crucial for the validity of the truncated ME, which is further illustrated in \cref{sec:optimizationDiscussion}. The optimized weights read
\begin{equation}
\lVert \hat{A}_t^\pm \rVert^2  =  \pm \overline{g_t} + \sqrt{\overline{g_t}^2 + V[g_t] + V[h_t]},
\label{eq:eigenvalues}
\end{equation}
with "variance" $V[x]=\overline{x^2} - \overline{x}^2$.  
Thus, the truncation is expected to provide a good approximation, as long as the variances of the real and imaginary parts of the bath correlation function are small. The truncated ME (\ref{eq:truncatedME}) becomes exact in the limit of a constant bath correlation function, i.e.\ energy independent, for which the variances vanish. This is also known as the singular coupling limit in which the bath correlation is time local $C_\tau = \alpha \delta(\tau)$ and the convolution operator $\hat{\mathbb{S}}_t=\alpha \hat{S}$ is proportional to the coupling operator with some real constant $\alpha$ \cite{GoriniKossakowski1976,PFPalmer1977}. In this limit the Lamb-shift vanishes, the optimal parameters reduce to $\lambda_t=|\alpha|^{1/2}$ and $\phi_t=0$, and only the positive jump operator $\hat{A}_t^+=(|\alpha|/2)^{1/2}\hat{S}$ contributes to the Redfield equation \footnote{For singular coupling with time local bath correlation $C_\tau = \alpha \delta(\tau)$ and convolution $\hat{\mathbb{S}}_t=\alpha \hat{S}$ with some real constant $\alpha$ the optimal parameters are most easily obtained from the basis independent form $\lambda^2_t=\lVert\hat{\mathbb{S}}_t\rVert/\lVert\hat{S}\rVert=\alpha$ and $\sin\varphi= \mathrm{Im}\tr_\mathrm{S}(\hat{S}\hat{\mathbb{S}}_t^\dagger)/\lVert\hat{S}\rVert\lVert\hat{\mathbb{S}}_t\rVert=\mathrm{Im}(\alpha)/\alpha=0$}. \newline
In order to estimate the quality of the approximation, let us have a look at the relative weight of the negative contribution. For this purpose, we will focus on Ohmic baths at inverse temperature $\beta$. Results for other bath models are presented in \cref{sec:generalbath}. The bath is characterized by the spectral density $J(\Delta)$ from which the bath correlation is obtained, $C_\tau=  \int_{-\infty}^{\infty} \exp[-i\Delta \tau/\hbar] J(\Delta)/(\exp[\beta\Delta]-1)\, d\Delta/\pi\hbar^2$. We consider $J(\Delta) = \gamma\Delta/(1+\Delta^2/E_c^2)$, with Drude cutoff at energy $E_c$, where the dimensionless factor $\gamma$ comprises the coupling strength relative to the energy scales of the system encoded in the level splittings $\Delta$ ta\-king values $\Delta_{kq}$. For this model the bath correlation $C_\tau$ is found to decay exponentially with time $\tau_B=\mathrm{max}\{\hbar/E_c, \hbar\beta/2\pi\}$ \cref{sec:bathcorrfunc}. Thus, assuming a large cutoff energy, the Markov approximation to the Redfield equation is valid if the coupling is small compared to the bath temperature. 
For com\-pu\-ting the long-time dynamics or the steady-state, one can replace $\hat{A}^\pm_t$ by $\hat{A}^\pm_\infty$ and obtains
\begin{equation}
\frac{\lVert \hat{A}_\infty^-\rVert^2}{\lVert \hat{A}_\infty^+\rVert^2} = \Big[\frac{1}{16}+\frac{\chi^2}{2}\Big] \beta^2\,V[\Delta] +O(\beta^4\overline{\Delta^4}, \overline{\Delta^2}/E_c^2),
\label{eq:tempscaling}
\end{equation}
where $\chi=\cot(\xi/2)/2 +\xi^2/\pi \sum_{l=1}^\infty \frac{1}{l(\xi^2 - (2\pi l)^2)}$ with $\xi=\beta E_c$. Note that \cref{eq:tempscaling} is found also for Ohmic baths with different cutoff (see \cref{sec:generalbath}). According to \cref{eq:tempscaling} the truncated negative term in the pseudo-Lindblad dissipator is small for a temperature that is large compared to the variance of the level splitting. Consequently  for sufficiently small $\beta$ the truncated ME (\ref{eq:truncatedME}) should be applicable beyond the zero coupling limit.
\section{Relation to Brownian motion} One of the few exactly solvable open quantum systems is the paradigmatic example of the damped harmonic oscillator \cite{HuPazZhang1992,KarrleinGrabert97,Paz94,GrabertWeiss84}. In the high-temperature regime it is described by the equation of Brownian motion, which is a Lindblad master equation \cite{breuerpetruccione}. However, there is no corresponding equation of motion for general systems. We now demonstrate that the truncated master equation reproduces the equation of Brownian motion in the high-temperature limit and, thus, might be seen as a generalization for it for general systems. \newline
The damped harmonic oscillator describes a particle with mass $M$ in a quadratic potential with oscillator frequency $\Omega$, whose position is coupled to a continuum of oscillator modes. The total system-bath Hamiltonian is given by
\begin{align}
	\hat{H}_\mathrm{tot} &= \frac{\hat{P}^2}{2M} + \frac{M \Omega^2}{2} \hat{Q}^2 \\ &+ \sum_k^\infty \Bigg[ \frac{\hat{p}_k^2}{2m_k} + \frac{m_k \omega_{k}^2}{2} \Big(\hat{q}_k - \frac{c_k}{m_k \omega_k^2} \hat{Q} \Big)^2 \Bigg],
	\label{eq:Htot}
\end{align}
with position $\hat{Q}$ and momentum $\hat{P}$ of the central oscillator. The coupling between system and bath is of the form $\hat{H}_\mathrm{SB} = \hat{Q} \otimes \hat{B}$ with bath operator $\hat{B}=\sum_k^\infty -c_k \hat{q}_k$ where the coefficients $c_k$ determine the coupling strength between the individual bath modes and the system. The model also takes into account the potential renormalization $H_\mathrm{RN} = \sum_k^\infty  c_k^2/(m_k \omega_k^2) \hat{Q}^2 = 2M h_\infty(0) \hat{Q}^2$, which cancels the damping kernel $h_\infty(0)$ in the imaginary part of the bath correlation function. The Redfield equation takes the form of \cref{eq:lambshift,eq:redfielddissipator} for which one identifies the dimensionless coupling operator $\hat{S}=1/\sqrt{2} (\hat{a} + \hat{a}^\dagger)$ and explicitly obtains the convolution $\hat{\mathbb{S}}_\infty=1/\sqrt{2}\, (G_\infty(-\Omega)\ \hat{a} + G_\infty(\Omega)\ \hat{a}^\dagger )$, where $\hat{a}^\dagger$ ($\hat{a}$) is the creation (annihilation) operator for eigenmodes of the central oscillator that is related to the position and momentum via $\hat{a} = (M\Omega/2\hbar)^{1/2} (\hat{Q} + (i/M\Omega) \hat{P})$. The detailed form of the bath correlation function depends on the particular bath model. However, in the high-temperature limit and by assuming a large cutoff energy for the spectral density one obtains the universal result
\begin{align}
	G_\infty(\Omega) \simeq \gamma \bigg[\frac{k_\mathrm{B} T}{\hbar} - i\ \chi \Omega \bigg],
\end{align}
where $k_\mathrm{B}$ is the Boltzmann constant and where $\chi$ is a real number that depends on how the cutoff is introduced. Generically the real part of the bath correlation function is given by the thermal time $\gamma k_\mathrm{B}T/\hbar$. For the imaginary part note that the potential renormalization cancels the damping kernel and in the limit of large cutoff energies the vacuum fluctuations decay such that only the antisymmetric thermal noise contributes. \newline
In order to construct the truncated master equation we calculate the parameters $\lambda^2_\infty$ and $\varphi_\infty$ by following the optimization procedure. For the damped harmonic oscillator one finds explicitly, $\lambda_\infty^2 =1/\sqrt{2}\, \sqrt{|G(\Omega)|^2 + |G(-\Omega)|^2}$, $\sin\varphi_\infty = \lVert \hat{S} \rVert/ (2\, \lVert \hat{\mathbb{S}}_\infty \rVert)\mathrm{Im} [G(\Omega) + G(-\Omega) ]$. In the high-temperature limit this reduces to $\lambda_\infty \simeq \sqrt{\gamma k_\mathrm{B} T/\hbar}$ and $\varphi_\infty\simeq0$. Finally we have everything at hand to write down the jump operator of the truncated master equation,
\begin{align}
	\begin{aligned}
		\hat{A}^+_\infty &= \sqrt{\frac{\gamma}{2}}  \bigg[\sqrt{\frac{k_\mathrm{B} T}{\hbar}} \hat{S} +  \frac{1}{\sqrt{k_\mathrm{B} T/\hbar}} \hat{\mathbb{S}}_t\bigg] \\
		&= \sqrt{\frac{\gamma \Omega}{2}}  \bigg[ \sqrt{\frac{4 M k_\mathrm{B} T}{\hbar^2}} \hat{Q} + \frac{1}{\sqrt{(1/\chi^2) M k_\mathrm{B} T}} \hat{P} \bigg].
	\end{aligned}
\end{align}
This is exactly the same jump operator as for the equation of Brownian motion \cite{breuerpetruccione}. 
\section{Concrete example} We further test our method for the extended Hubbard chain with $N$ spinless fermions and $l$ sites, described by
\begin{equation}
	\hat{H}_\mathrm{ S } = -J \sum\limits_{i=1}^{l-1} \left( \hat{a}_i^\dagger \hat{a}_{i+1} + \hat{a}_{i+1}^\dagger \hat{a}_i \right) + V \sum\limits_{i=1}^{l-1} \hat{n}_i  \hat{n}_{i+1},
	\label{eq:fermihubbard}
\end{equation}
\begin{figure}[b]
	\includegraphics[width=1\columnwidth]{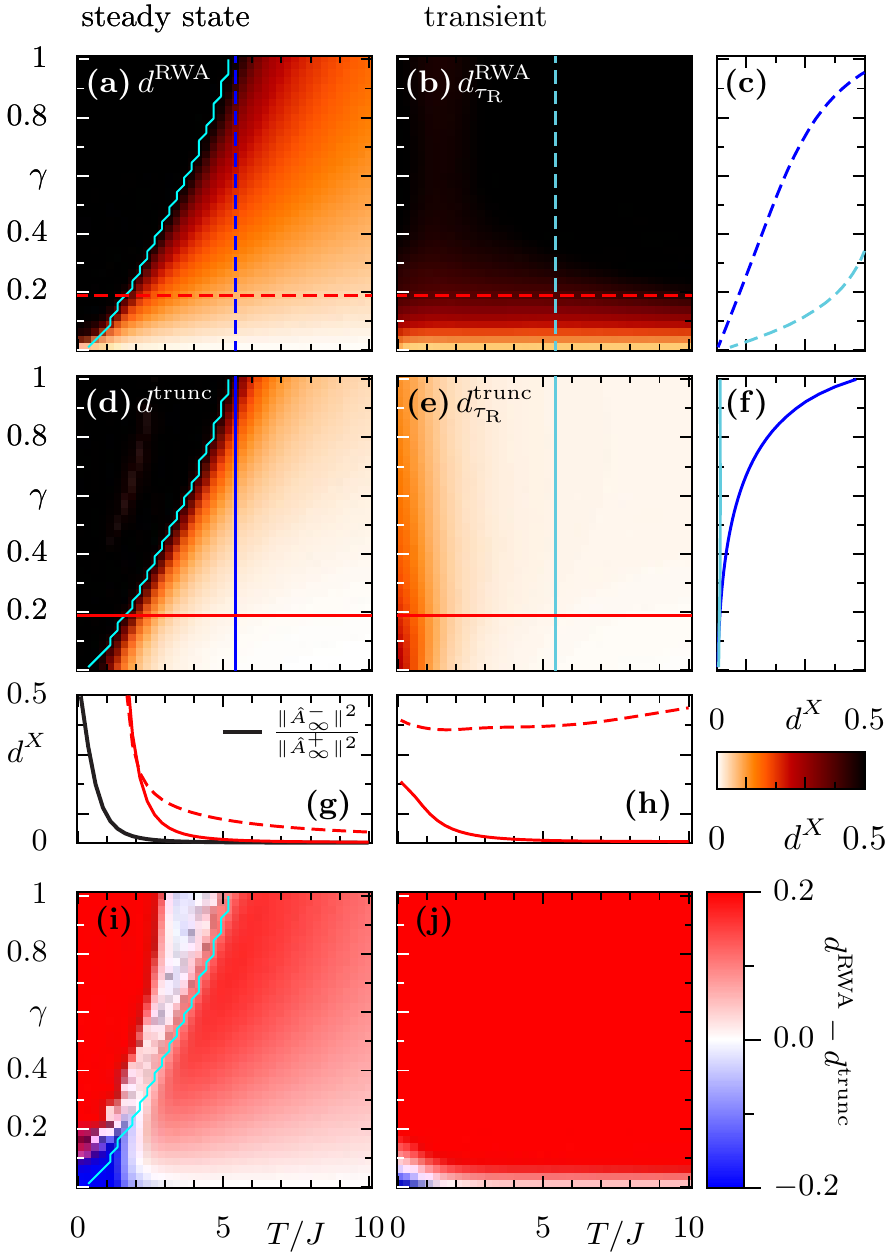}
	\caption{Error $d^X$ of the RWA [(a),(b), dashed lines] and of the truncated ME [(d),(e), solid lines] as a function of the bath temperature $T/J$ and coupling strength $\gamma$ for the steady-state (left panels) and for the transient with averaging time $\tau_\mathrm{R}=2\hbar/\gamma J$ (right panels). The smaller panels show cuts for fixed $\gamma=0.19$ [(g), (h), along the vertical blue lines in (a,b,d,e)] and fixed $T/J=5.43$ [(c), (f), along the horizontal red lines in (a,b,d,e)]. To the left of the wiggly cyan line in (a),(d),(i) the Redfield steady-state acquires negative populations. The lower panels show $d^\mathrm{RWA}-d^\mathrm{trunc}$ for the steady and the transient state in (i) and (j), respectively. The parameters are $l = 8$, $N=4$, $V=2\,J$, $E_c=17\,J$.}
	\label{fig:distmeas}
\end{figure}
with annihilation and number operators $\hat{a}_i$ and $\hat{n}_i=\hat{a}^\dagger_i\hat{a}_i$ at site $i$. The tunneling parameter $J$ quantifies the kinetic energy of the particles and $V$ is the interaction energy of particles occupying adjacent sites. The system is driven by a local heat bath at temperature $T$ that couples to the density $\hat{n}_1$. For a bath that couples globally to all sites similar results are found as outlined in the section below. In order to quantify the deviation of the RWA and truncation approach from the Redfield result, we introduce the error measure \cite{dodonovmanko,gilchristlangford},
\begin{align}
d^\mathrm{RWA/trunc} = \frac{1}{2} \mathrm{tr} \sqrt{(\hat{\rho}^\mathrm{RWA/trunc} - \hat{\rho}^\mathrm{Red})^2} \in [0,1]. \label{eq:distmeas}
\end{align}
\subsection{Steady-state} In equilibrium the total system-bath compound thermalizes at the given temperature and by tracing over the bath degrees of freedom the reduced density matrix of the system has the generalized Gibbs form, $\rho_{th}=\tr_B \exp[-\beta \hat{H}_\mathrm{tot}]/\tr_\mathrm{S}\tr_B \exp[-\beta \hat{H}_\mathrm{tot}] $ \cite{TMori2008}. However, there is yet no ME that gives this steady-state solution in all orders of the coupling strength $\gamma$. The RWA only captures the zeroth order contribution, whereas the Redfield equation also correctly reproduces the coherences in first order \cite{JThingaPHaenggi2012,breuerpetruccione,VRomero1989,EGevaERosenman2000}. We will now compare the steady-state errors defined by \cref{eq:distmeas} for the truncation method with those of the RWA. Both are plotted versus temperature $T/J$ and coupling strength $\gamma$ in \cref{fig:distmeas} (a) and (d). For fixed temperature $T/J=5.43$ in the weak coupling regime the error of the RWA scales linear with $\gamma$ [\cref{fig:distmeas} (c) dashed dark], whereas the error of the truncation is smaller and of higher order [\cref{fig:distmeas} (f) solid dark]. Also the result of the truncated ME shows good agreement for large temperatures. For fixed coupling strength $\gamma=0.19$ in \cref{fig:distmeas} (g) we can see that $d_\mathrm{trunc}$ (solid red line) decays rapidly with temperature, like $\Vert\hat{A}^-_\infty\rVert/^2\Vert\hat{A}^+_\infty\rVert^2$ (black solid line), whereas $d_\mathrm{RWA}$ decays much slower. From $d_\mathrm{RWA}-d_\mathrm{trunc}$ in \cref{fig:distmeas} (i), it is evident that the steady-state solution of the truncated ME is in better agreement with the Redfield result than the RWA for all parameters except for very weak coupling and low temperatures. Note that this is also the regime, in which the Redfield steady-state acquires unphysical negative probabilities as marked by the bright cyan line in Figs.\ \ref{fig:distmeas} (a), (d) and (i). This is a known problem of the Redfield formalism \cite{VRomero1989,ASuarezIOppenheim92,PPechukas94,EGevaERosenmann2000}. Namely, for low temperatures the bath correlation time becomes large compared to the coupling strength for which the Born-Markov approximation no longer holds~\cite{deVaga2017}. This is in accordance with our analysis that for low temperatures the weight of the negative contribution in the pseudo-Lindblad dissipator grows significantly causing the Redfield steady-state to have negative populations. Just recently it has been argued that this loss of positivity indicates the failure of the weak coupling assumption \cite{Strunz20}.
\subsection{Transient dynamics} Let us now study the relaxation dynamics starting from the system's ground state. We evaluate the error for the transient dynamics by introducing the time averaged distance measure $d_{\tau_\mathrm{R}}^\mathrm{RWA/trunc}=(1/\tau_\mathrm{R})\int_0^{\tau_\mathrm{R}} d^\mathrm{RWA/trunc}(t)\ dt$,
where we obtain the solutions $\hat{\rho}^{X}(t)$ by direct integration of the particular ME.
We aim at choosing $\tau_\mathrm{R}$ big enough to cover the transient regime but small enough not to capture the steady-state properties. For the parameters discussed here $\tau_\mathrm{R}=2\hbar/\gamma J$ turns out to be a reasonable choice. \newline
The RWA provides a poor prediction of the transient dynamics [\cref{fig:distmeas} (b)]. A large error of $0.5$ (the maximum value plotted) is reached already for very small coupling $\gamma\simeq0.3$ [\cref{fig:distmeas} (c) cyan dashed line]. For short times the neglect of non-resonant terms with $\Delta_{qk}\ne\Delta_{q'k'}$ in the RWA overestimates the relaxation \cite{BBalzerGStock2004}. Here the truncation method [\cref{fig:distmeas} (e)] clearly outperforms the RWA. For all parameters except a small regime for $T/J\le1$ and $\gamma\le0.07$ the time averaged error for the truncated ME is not only smaller than the one in the RWA [\cref{fig:distmeas} (j)] but also very close to zero [\cref{fig:distmeas} (f) solid bright, (h) solid red]. 
\subsection{Globally coupled bath}
The coupling operator $\hat{S}$ of the system-bath interaction $\hat{H}_\mathrm{SB}$ defines in which way the bath is coupled to the system. Local coupling operators are most relevant for transport properties, where the baths couple to the edges of a system. In this section we briefly discuss the case when a bath couples globally to the system. 
\begin{figure}[b]
	\includegraphics{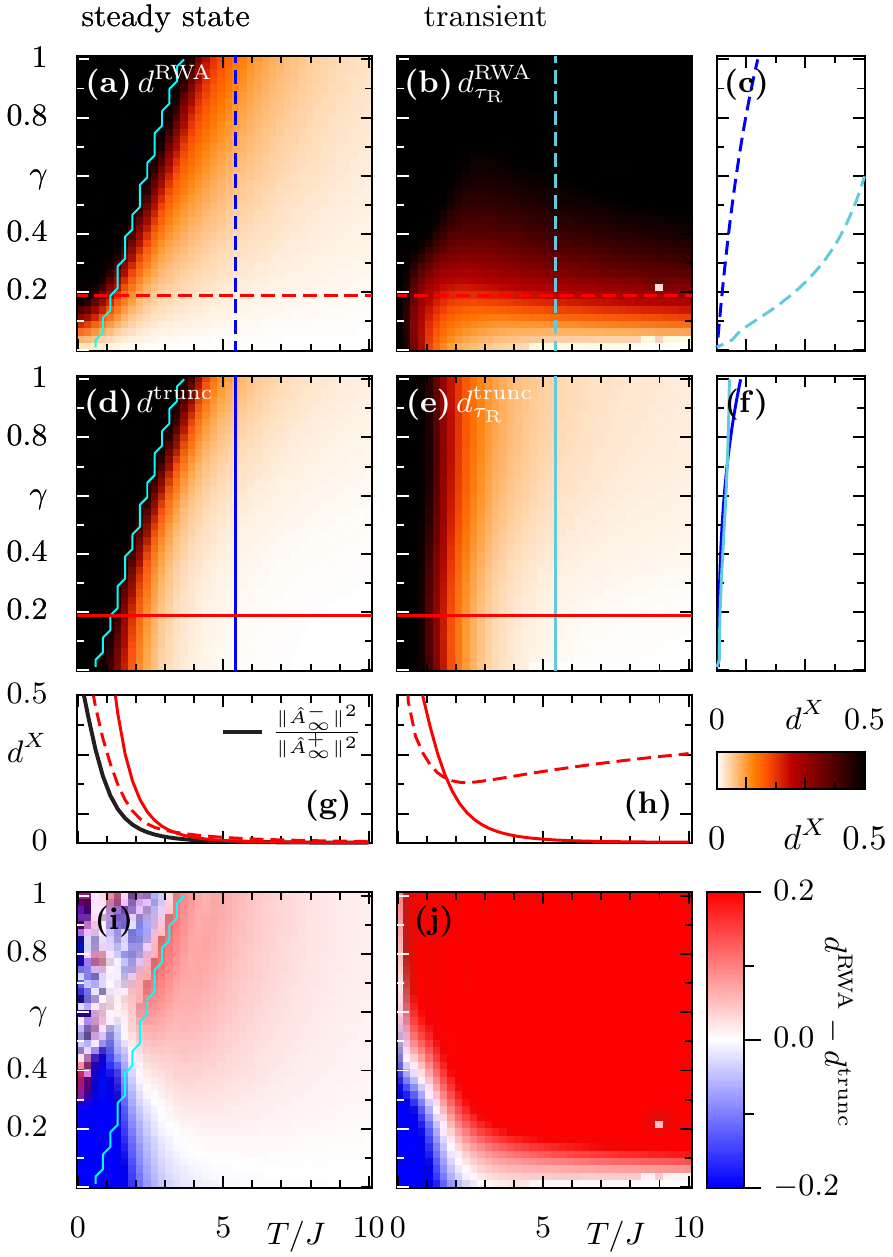}
	\caption{For a bath that couples globally error $d^X$ of the RWA [(a),(b), dashed lines] and of the truncated master equation [(d),(e), solid lines] as a function of the bath temperature $T/J$ and system-bath coupling strength $\gamma$ for the steady-state (left panels) and for the transient with averaging time $\tau_\mathrm{R}=1\hbar/\gamma J$ (right panels). The smaller panels show cuts for fixed $\gamma=0.19$ [(g), (h), along the vertical blue lines in (a,b,d,e)] and fixed $T/J=5.43$ [(c), (f), along the horizontal red lines in (a,b,d,e)]. To the left of the wiggly cyan line in (a),(d),(i) the Redfield steady-state acquires negative populations. The lower panels show $d^\mathrm{RWA}-d^\mathrm{trunc}$ for the steady and the transient state in (i) and (j), respectively. The parameters are $l = 8$, $N=4$, $V=2\,J$, $E_c=40\,J$.}
	\label{fig:errglob}
\end{figure}
For models where the coupling operator itself is a global quantity, e.g.\ for the damped harmonic oscillator, all the previous expressions hold. However, in particular for the extended Hubbard model studied in the main text the coupling operator $\hat{S} = \sum_{i=1}^l \hat{n}_i= N$ is not a reasonable choice, since it simply corresponds to the total particle number which is conserved. Instead one has to consider a system-bath interaction Hamiltonian that consists of several coupling terms. We choose $\hat{S}_\alpha = \hat{n}_\alpha$ where the index $\alpha$ labels independent baths of the same temperature. In \cref{fig:errglob} we repeat the analysis of \cref{fig:distmeas} of the main text but for a bath that couples globally to the system rather than to a single site only. Here the trace distance to the full Redfield result in \cref{eq:distmeas} again serves as an error measure for the RWA and the truncated master equation, respectively.  As compared to a bath that only couples locally, here the damping dominates the coherent dynamics. The relaxation time becomes shorter and therefore we compute the time averaged distance measure in \cref{fig:errglob} (b), (e), (h) and (j) for $\tau_\mathrm{R}=1\hbar/\gamma J$ (as compared to $\tau_\mathrm{R}=2\hbar/\gamma J$, which was used for the local bath). The initial state is a coherent superposition of the ground state and first the excited state. \newline
By looking at the relative error $d^\mathrm{RWA} - d^\mathrm{trunc}$ for the steady state in \cref{fig:errglob} (i) and for the dynamics in \cref{fig:errglob} (j), it is evident that  for weak coupling and low temperature the RWA performs better, whereas for finite coupling and higher bath-temperature the truncated master equation is favourable. All in all the case of a global bath is qualitatively similar to the case of a local bath. 
\section{Nonequilibrium steady-state} Finally, we examine properties of the nonequilibrium steady-state of the driven-dissipative system, focusing on parameters, where the RWA is known to be an inadequate description \cite{HWichterich2007}. The system is driven by two local baths at different temperature $T_L<T_R$ that couple to the occupations $\hat{n}_1$ and $\hat{n}_l$ of the outermost sites of the chain, respectively. 
\begin{figure}[b]
	\centering
	\includegraphics[width=1\columnwidth]{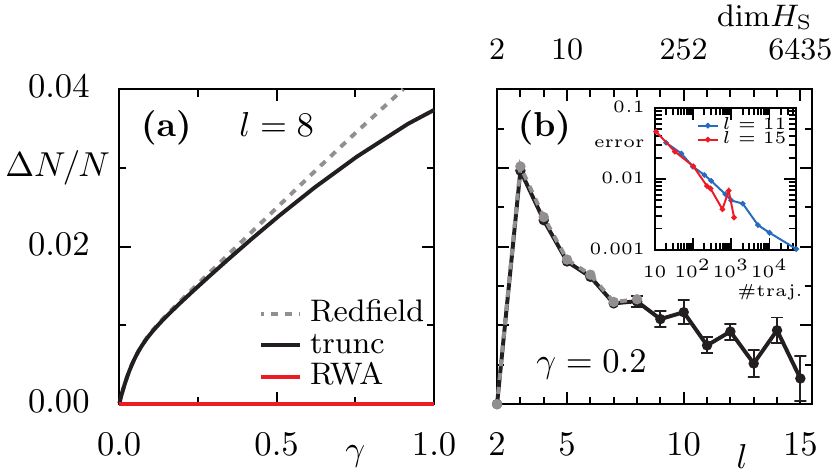}
	\caption{Particle imbalance in nonequilibrium steady-state for $N=\lfloor l/2\rfloor$ $U=2J$, $E_c=17J$, $T_L=7J$, $T_R=13J$. Plotted (a) for $l=8$ versus $\gamma$ and (b) for $\gamma =0.2$ versus $l$ and the Hilbert space dimension $\mathrm{dim}H_\mathrm{S}$. For system sizes $l\le8$ it is calculated via sparse LU decomposition (Redfield in dashed grey, truncated ME in solid black, RWA solid red). For $l\ge8$ the truncated ME is solved by quantum trajectory simulations. The inset in (b) shows the statistical error as a function of the number of trajectories for $l=11$ and $l=15$. We average over $5\cdot10^4$, $10^4$, $6\cdot10^3$ and $10^3$ trajectories for $l=8,9,10,11$, $l=12,13$, $l=14$ and $l=15$, respectively. Lines are guides to the eye.}
	\label{fig:particleimbalance}
\end{figure}
In \cref{fig:particleimbalance} the particle imbalance $\Delta N= N_L-N_R$ in the nonequilibrium steady-state is shown, where $N_L=\sum_{i<l/2} \expval{\hat{n}_i}$ and $N_R=\sum_{i>l/2} \expval{\hat{n}_i}$ count the particles on the left and right half of the chain, respectively. \newline
According to the thermoelectric effect \cite{thermoelectric} a greater particle mobility near to the hotter, right reservoir is expected such that the particle density tends to the left side of the chain, i.e.\ $\Delta N>0$. However, this is not captured by the RWA. Just as in equilibrium the off-diagonal elements of the density matrix decay and the steady-state is diagonal in the eigenbasis of $\hat{H}_\mathrm{S}$. Since the eigenstates reflect the symmetry of the system that has no preferred orientation the nonequilibrium steady-state in RWA localizes evenly among the left and right half of the chain [\cref{fig:particleimbalance} (a) solid red]. \newline
For finite coupling parity is broken and finite off-diagonal matrix elements of the nonequilibrium steady-state give a non-zero contribution to particle imbalance. This is well captured by the truncated ME [\cref{fig:particleimbalance} (a)]. Furthermore its Lindblad-form allows the use of quantum trajectory simulations \cite{Dalibard92}. This is beneficial especially for many body systems for which the Hilbert space dimension grows exponentially with the system size. Thus, the truncated ME allows to study larger systems that are hardly accessible by direct integration of the Redfield equation [\cref{fig:particleimbalance} solid black].
\section{Conclusion}We have derived an alternative Lindbladian approximation to the Redfield ME. It provides an accurate description in large parameter regimes, where the RWA fails, in particular for non-equilibrium scenarios like transient dynamics and non-equilibrium steady states which are non-trivial also in the high-temperature regime. It, thus, allows for efficient quantum trajectory simulations also beyond ultra-weak coupling. 
\begin{acknowledgments}
This research was funded by the Deutsche Forschungsgemeinschaft (DFG) via the Research Unit FOR 2414 under the Project No. 277974659. We thank Daniel Vorberg and Roland Ketzmerick for helpful discussions in the early stage of this project. We thank the developers of QuTiP \cite{qutip}, which was used for numerical calculations.
\end{acknowledgments}

\appendix
\section{Bath correlation function} \label{sec:bathcorrfunc} Generally for open quantum systems the bath model is defined by the bath Hamiltonian $\hat{H}_\mathrm{B}$ and the system-bath coupling Hamiltonian $\hat{H}_\mathrm{SB}=\hat{S}\otimes\hat{B}$. In the Redfield master equation (\ref{eq:redfield}) in the main text the details of the bath are incorporated via the bath correlation $C_\tau=\tr_\mathrm{B}(\hat{B}_\tau\hat{B}\hat{\rho}_\mathrm{B})/\hbar^2$, with $\hat{B}_{\tau}=\exp[i\hat{H}_\mathrm{B} \tau/\hbar] \hat{B}\exp[-i \hat{H}_\mathrm{B}\tau/\hbar]$. We consider a thermal bath $\hat{\rho}_\mathrm{B}=\exp[-\beta\hat{H}_\mathrm{B}]/Z_\mathrm{B}$ at inverse temperature $\beta$. The bath correlation then assumes the form~\cite{JThingaPHaenggi2012},
\begin{align}
	C_\tau =  \int\limits_{-\infty}^{\infty} e^{i\Delta\tau/\hbar}\, \frac{J(\Delta)/\hbar}{e^{\beta\Delta}-1}\, \frac{d\Delta}{\pi\hbar},
\end{align}
where the bath model is specified by means of the antisymmetric spectral density $J(\Delta)=-J(-\Delta)$. We consider an Ohmic bath with Drude cutoff at energy $E_c$,
\begin{align}
	J(\Delta) = \frac{\gamma \Delta}{1+(\Delta/E_c)^2}.
	\label{eq:ohmic}
\end{align}
In the upper complex plane the integrand decays exponentially such that the integral can be solved by the residue theorem. The Drude spectral density becomes singular at the complex cutoff energy $i E_c$ for which the residue is $\mathrm{Res}(J(\Delta), \Delta=i E_c)=\gamma E_c^2/2$. The Bose function has poles at the complex Matsubara energies $\nu_l=2\pi l/\beta$, which is seen by noting that $[\exp(\beta\Delta)-1]^{-1}=(1/2)[\coth(\beta\Delta/2) - 1]$ and by making use of the series expansion, $\coth(\beta\Delta/2)=2/(\beta\Delta) \sum_{l=-\infty}^{\infty} 1/(1+\nu_l^2/\Delta^2)$. The residues are given by $\mathrm{Res}([\exp(\beta\Delta)-1]^{-1}, \Delta=i \nu_l)=1/\beta$. Altogether the bath correlation reads
\begin{align}
	C_\tau = &\frac{\gamma E_c^2}{2\hbar^2}\,[\cot(\beta E_c/2) - i]\, e^{-E_c\tau/\hbar} \notag \\& - \frac{2\gamma}{\hbar^2\beta} \sum\limits_{l=1}^\infty \frac{\nu_l\, e^{-\nu_l\tau/\hbar}}{1 - (\nu_l/E_c)^2}.
\end{align}
We also introduce the bath correlation function,
\begin{align}
	G_t(\Delta) = g_t(\Delta) + i h_t(\Delta) = \int\limits_0^t e^{-i\Delta\tau/\hbar}\,C_\tau\,d\tau,
\end{align}
with real valued $g_t$, $h_t$ which is advantageous for the energy-basis representation of the RWA and which is used to determine the optimal values for $\lambda$ and $\varphi$ in the pseudo-Lindblad dissipator. Since the time dependence only arises in the exponentials this integral can be carried out straightforwardly. In the long-time limit the real part simplifies to
\begin{align}
	g_\infty(\Delta) = \frac{J(\Delta)/\hbar}{e^{\beta\Delta}-1},
\end{align}
which implies detailed balance within the RWA \cite{breuerpetruccione,RAlicki2007}. The imaginary part consists of three parts,
\begin{widetext}
\begin{align}
	h_\infty(\Delta)=  \underbrace{\frac{-\gamma E_c}{2\hbar}}_{h_\infty(0)} + \underbrace{\frac{\gamma\Delta^2 E_c}{2\hbar (E_c^2 + \Delta^2)}}_{ h_\infty^\mathrm{vac}(\Delta)} + \underbrace{(\Delta/\hbar) \gamma \Bigg[ \frac{-E_c^2}{2(E_c^2 + \Delta^2)} \cot(\beta E_c/2) +  + \frac{2}{\beta}\,\sum\limits_{l=1}^\infty \frac{\nu_l}{(\Delta^2+\nu_l^2)(1-\nu_l^2/E_c^2)} \Bigg]}_{h_\infty^\mathrm{th}(\Delta)},
	\label{eq:dampvacth}
\end{align}
\end{widetext}
the damping kernel  $h_\infty(0)$, the temperature independent and symmetric part $h_\infty^\mathrm{vac}(\Delta)$, which describes vacuum fluctuations, and the antisymmetric part $h_\infty^\mathrm{th}(\Delta)$, which describes thermal noise.
\section{Pseudo-Lindblad equation} \label{sec:pseudoLindblad} The dynamics of the reduced density matrix ${\hat{\rho}=\tr_\mathrm{B}(\hat{\rho}_\mathrm{tot})}$ is described by the Redfield equation $\dot{\hat{\rho}} = (-i/\hbar) [\hat{H}_\mathrm{S} + \hat{H}_t^\mathrm{LS}, \hat{\rho}] + \mathcal{D}_t^\mathrm{Red}[\hat{\rho}]$ with $ \hat{H}_t^\mathrm{LS}$ and $\mathcal{D}_t^\mathrm{Red}$ given in \cref{eq:lambshift,eq:redfielddissipator}
with the system operator $\hat{S}$ and the convolution with the bath correlation $\hat{\mathbb{S}}_t = \int_0^{t} C_\tau\ \hat{S}_{-\tau}\ d\tau$. Here we derive the pseudo-Lindblad representation of the Redfield dissipator in \cref{eq:pseudolindblad}, 
where we have introduced the new jump operators $\hat{A}_t^{\pm}$ in \cref{eq:jumpoperators},
with $\lambda_t^\pm=\lambda_t e^{\mp i \frac{\varphi_t}{2}}$  and arbitrary real, time-dependent parameters $\lambda_t$ and $\varphi_t$. Essentially the symmetrized and antisymmetrized combination ensures that only the off-diagonal terms $\hat{S} \hat{\mathbb{S}}_t^\dagger$ and $\hat{\mathbb{S}}_t \hat{S}$ survive and the diagonals  $\hat{\mathbb{S}}_t\hat{\mathbb{S}}_t^\dagger$ and $\hat{S} \hat{S}$ cancel. For the first two terms in the dissipator we have,
\begin{widetext}
\begin{alignat}{3}
	\hat{A}_t^+ \hat{\rho} \hat{A}_t^{+ \dagger}  - \hat{A}_t^- \hat{\rho} \hat{A}_t^{- \dagger}  =& \frac{1}{2 \cos \varphi_t} && \bigg[ \Big(\lambda_t^+ \hat{S} + \frac{1}{\lambda_t^+} \hat{\mathbb{S}}_t\Big) \hat{\rho} \Big(\lambda_t^{+*} \hat{S} + \frac{1}{\lambda_t^{+*}} \hat{\mathbb{S}}_t^\dagger \Big) - \Big(\lambda_T^- \hat{S} - \frac{1}{\lambda_t^-} \hat{\mathbb{S}}_t\Big) \hat{\rho} \Big(\lambda_t^{-*} \hat{S} - \frac{1}{\lambda_t^{-*}} \hat{\mathbb{S}}_t^\dagger\Big) \bigg], \notag \\ 
	= &\frac{1}{2 \cos \varphi_t} && \bigg[ \Big(|\lambda_t^+|^2 - |\lambda_t^-|^2\Big) \hat{S}\hat{\rho}\hat{S} + \Big(\frac{1}{|\lambda_t^+|^2} - \frac{1}{|\lambda_t^-|^2}\Big) \hat{\mathbb{S}}_t \hat{\rho}\hat{\mathbb{S}}_t^\dagger \bigg] + \notag \\
	&+\frac{1}{2 \cos \varphi_t}& & \bigg[  \Big(\frac{\lambda_t^+}{\lambda_t^{+*}} + \frac{\lambda_t^-}{\lambda_t^{-*}}\Big)    \hat{S} \hat{\rho}  \hat{\mathbb{S}}_t^\dagger + \Big(\frac{\lambda_t^{+*}}{\lambda_t^+} + \frac{\lambda_t^{-*}}{\lambda_t^-}\Big) \hat{\mathbb{S}}_t \hat{\rho} \hat{S}  \bigg],
\end{alignat}
\end{widetext}
where the diagonal terms in the second line cancel due to $|\lambda_t^\pm|^2=\lambda_t^2$. By noting $\frac{\lambda_t^{+*}}{\lambda_t^+} + \frac{\lambda_t^{-*}}{\lambda_t^-}=e^{-i\varphi_t}+e^{i\varphi_t}=2\cos \varphi_t$ in the off-diagonal terms in the third line, we see that both the absolute value $\lambda_t$ and the phase $\varphi_t$ cancel in the pseudo-Lindblad equation. Finally, one arrives at, 
\begin{align}
	\hat{A}_t^+ \hat{\rho} \hat{A}_t^{+ \dagger}  - \hat{A}_t^- \hat{\rho} \hat{A}_t^{- \dagger}   =  \hat{S} \hat{\rho} \hat{\mathbb{S}}_t^\dagger + \hat{\mathbb{S}}_t \hat{\rho}\hat{S},
\end{align}
which is the first part of the Redfield dissipator \cref{eq:redfielddissipator} in the main text. Likewise we can show that the remaining terms follow analogously, where only the off-diagonal term $\hat{S} \hat{\mathbb{S}}_t^\dagger$  and its hermitian conjugated survive and the parameters $\lambda_t$ and $\varphi_t$ cancel out. 
\section{Optimal choice for $\lambda_t$ and $\varphi_t$} \label{sec:optimization} Since the pseudo-Lindblad equation is an exact representation of the Redfield equation, it does not depend on the choice of $\lambda_t$ and $\varphi_t$. However, these parameters change the relative weight of the negative contribution, and thus have an influence on the truncated master equation. Here we find the optimal values to minimize the weight of the negative contribution and thus to truncate it. \newline 
Thereby we calculate the weight with the Frobenius norm of the jump operators $\lVert \hat{A}_t^\pm \rVert^2 = \mathrm{tr}_\mathrm{S}(\hat{A}_t^\pm \hat{A}_t^{\pm\dagger})$. By making use of the additivity identity $\lVert \hat{A}+ \hat{B}\rVert^2= \lVert \hat{A}\rVert^2 + 2\mathrm{Re}\tr_\mathrm{S}(\hat{A}\hat{B}^\dagger) + \lVert\hat{B}\rVert^2$, which holds for any operators $\hat{A}$ and $\hat{B}$, the weights read
\begin{align}
	\lVert \hat{A}_t^\pm \rVert ^2  = \frac{1}{2 \cos \varphi_t} \Big[&  \lambda_t^2 \lVert\hat{S}\rVert^2 \pm 2 \cos\varphi_t \mathrm{Re}\tr_\mathrm{S}(\hat{S}\hat{\mathbb{S}}_t^\dagger) - \notag \\ &2 \sin\varphi_t\mathrm{Im}\tr_\mathrm{S}(\hat{S}\hat{\mathbb{S}}_t^\dagger) + \frac{1}{\lambda_t^2} \lVert\hat{\mathbb{S}}_t\rVert^2 \Big].
	\label{eq:weight}
\end{align}
In the following the weight of the negative contribution $\lVert \hat{A}_t^-\rVert^2$ is minimized by varying $\lambda_t^2$ and $\varphi_t$ at fixed but arbitrary time $t$. To make this clear the index is dropped in the subsequent discussion. We begin with the variation with respect to $\lambda^2$. The necessary condition for a minimum is,
\begin{align}
	0 = \left.\frac{\partial}{\partial \lambda^2} \lVert \hat{A}^-\rVert^2\right\vert_{\lambda_\mathrm{opt}}  =  \frac{1}{2 \cos \varphi} \Big[ \lambda_\mathrm{opt}^2 \lVert\hat{S}\rVert^2- \frac{1}{\lambda_\mathrm{opt}^2} \lVert\hat{\mathbb{S}}\rVert^2\Big],
	\label{eq:optlambda}
\end{align}
from which the optimal parameter $\lambda_\mathrm{opt}^2=\lVert\hat{\mathbb{S}}\rVert/\lVert\hat{S}\rVert$ is deduced. Furthermore the variation with respect to the complex angle $\varphi$ gives,
\begin{align}
	0 = \left.\frac{\partial}{\partial\varphi} \lVert \hat{A}^-\rVert^2\right\vert_{\substack{\varphi_\mathrm{opt}\\ \lambda_\mathrm{opt}}} = \frac{\lVert\hat{S}\rVert\lVert\hat{\mathbb{S}}\rVert \sin\varphi_\mathrm{opt} - \mathrm{Im}\tr_\mathrm{S}(\hat{S}\hat{\mathbb{S}}^\dagger) }{(\cos \varphi_\mathrm{opt})^2},
	\label{eq:optphi}
\end{align} 
and leads to the optimal choice of $\sin\varphi_\mathrm{opt}= \mathrm{Im}\tr_\mathrm{S}(\hat{S}\hat{\mathbb{S}}^\dagger)/\lVert\hat{S}\rVert\lVert\hat{\mathbb{S}}\rVert$. The same results are obtained by varying with respect to $\varphi$ first and with respect to $\lambda$ second. For the sake of completeness one straightforwardly verifies the sufficient conditions for local minima $\left.\frac{\partial^2}{\partial x^2} \lVert \hat{A}^-\rVert^2\right\vert_{\substack{\varphi_\mathrm{opt}\\ \lambda_\mathrm{opt}}} > 0$ with $x=\lambda^2$ and $x=\varphi$. It turns out that the extremal condition also holds for the ratios,
\begin{align}
	\begin{aligned}
		0 &= \left.\frac{\partial}{\partial \lambda^2} \frac{\lVert \hat{A}^-\rVert^2}{\lVert \hat{A}^+\rVert^2}\right\vert_{\lambda_\mathrm{opt}}  \\ &= \frac{4\cos\varphi \mathrm{Re}\tr_\mathrm{S}(\hat{S}\hat{\mathbb{S}}_t^\dagger)}{\lVert\hat{A}^+\rVert^4} \Big[ \lambda_\mathrm{opt}^4 \lVert\hat{S}\rVert^2-\lVert\hat{\mathbb{S}}^\dagger\rVert^2 \Big], \\
		0 &= \left.\frac{\partial}{\partial\varphi}\frac{\lVert \hat{A}^-\rVert^2}{\lVert \hat{A}^+\rVert^2}\right\vert_{\substack{\varphi_\mathrm{opt}\\ \lambda_\mathrm{opt}}}  \\ &=\frac{2\lambda_\mathrm{opt}^2 \mathrm{Re}\tr_\mathrm{S}(\hat{S}\hat{\mathbb{S}}^\dagger)}{\lVert\hat{A}^+\rVert^4} \Big[ \lVert\hat{\mathbb{S}}\rVert^2 \sin\varphi_\mathrm{opt}- \frac{\lVert\hat{\mathbb{S}}\Vert}{\lVert\hat{S}\rVert}\mathrm{Im}\tr_\mathrm{S}(\hat{S}\hat{\mathbb{S}}^\dagger) \Big],
	\end{aligned}
\end{align}
such that the optimal parameters minimize the weight of the negative contribution both absolutely and relatively to the positive one. By reintroducing the time dependence of the convolution operator $\hat{\mathbb{S}}_t$ for the optimal parameters the weights take the values
\begin{align}
	\lVert \hat{A}_t^\pm\rVert^2 = \pm \mathrm{Re}\tr_\mathrm{S}(\hat{S}\hat{\mathbb{S}}_t^\dagger) + \sqrt{\lVert\hat{S}\rVert^2 \lVert\hat{\mathbb{S}}_t\rVert^2 - [\mathrm{Im}\tr_\mathrm{S}(\hat{S}\hat{\mathbb{S}}_t^\dagger)]^2}.
\end{align}
To get an explicit form for a bath model the trace is performed in the eigenbasis of the system,
\begin{align}
	\tr_\mathrm{S}(\hat{S}\hat{\mathbb{S}}_t^\dagger) &= \sum\limits_{qk} \matrixelement{q}{\hat{S}}{k} \matrixelement{k}{\hat{\mathbb{S}}_t^\dagger}{q} = \sum\limits_{qk} |S_{qk}|^2 G_t^*(\Delta_{qk}), \label{eq:trSS} \\
	\lVert\hat{\mathbb{S}}_t\rVert^2 &= \sum\limits_{qk} \matrixelement{q}{\hat{\mathbb{S}}_t}{k} \matrixelement{k}{\hat{\mathbb{S}}_t^\dagger}{q} = \sum\limits_{qk} |S_{qk}|^2 |G_t(\Delta_{qk})|^2,
\end{align}
where $G_t(\Delta)=g_t(\Delta) + i h_t(\Delta)$ is the bath correlation function, which is connected to $C_\tau$ via the integral $G_t(\Delta)=\int_0^t \exp[-i\Delta\tau/\hbar]\, C_\tau\, d\tau$. Without loss of generality the coupling matrix $S_{qk}=\matrixelement{q}{\hat{S}}{k}$ is assumed to be normed, i.e.\, $\sum_q |S_{qq}|^2=1$. The optimal parameters are given by $\lambda_t^4=\overline{g_t^2} + \overline{h_t^2}$ and $\sin\varphi_t=\overline{h_t}/(\overline{g_t^2} + \overline{h_t^2})^{1/2}$, where the overline denotes an average defined by $\overline{x} =\sum_{qk} x(\Delta_{qk}) |S_{qk}|^2$. Here $|S_{qk}|^2$ plays the role of a probability distribution. Finally the weights $\lVert\hat{A}_t^\pm\rVert^2$ are further simplified to \cref{eq:eigenvalues} in the main text, where $V[x]=\overline{x^2}-\overline{x}^2$ defines the "variance".
\section{Relative weight of the negative contribution} \label{sec:generalbath} In the following we compute the weights in the time independent pseudo-Lindblad equation and find the temperature scaling for an arbitrary spectral density $J(\Delta)$. The real part of the bath correlation function is  found to be
\begin{align}
	g_\infty(\Delta) &= \frac{J(\Delta)/\hbar}{e^{\beta\Delta}-1}=\frac{J(\Delta)}{2\hbar}\,[\coth(\beta\Delta/2) - 1]
\end{align}
where $\beta$ is the inverse bath temperature. Note that only averages that are symmetric in $\Delta$ contribute and that the spectral density is antisymmetric \cite{DVorbergAEckardt2015}. As a result the relative weight of the negative contribution scales only in even powers of the inverse bath temperature. In the high-temperature regime the weights reduce to
\begin{widetext}
\begin{align}
	\lVert\hat{A}_\infty^\pm\rVert^2 = \frac{1}{\hbar\beta}\Bigg\{\pm\Big[ \overline{\frac{J(\Delta)}{\Delta} }+ \beta^2\, \overline{\frac{J(\Delta)\Delta}{12}}\Big] + \sqrt{\overline{\frac{J(\Delta)^2}{\Delta^2}} + \beta^2\,\frac{5}{12}\, \overline{J(\Delta)^2} + \beta^2\, V[h_\infty]} + O(\beta^4\overline{\Delta^4})\Bigg\}.
\end{align}
\end{widetext}
If we assume $V[h_\infty]=O(\beta^2)$, which will be discussed for the Drude bath in more detail, the imaginary part of the bath correlation function only contributes in second order. In zeroth order the relevant expressions are  $\lVert\hat{A}_\infty^\pm\rVert^2\simeq (1/\hbar\beta) \{\pm \overline{J(\Delta)/\Delta} + \overline{J(\Delta)^2/\Delta^2}\}$ and the ratio of the weights becomes
\begin{align}
	\frac{\lVert \hat{A}_\infty^-\rVert^2}{\lVert \hat{A}_\infty^+\rVert^2} = \frac
	{ 1-\frac{\overline{(J(\Delta)/\Delta)^2}}{\overline{J(\Delta)/\Delta}} } 
	{1 +\frac{\overline{(J(\Delta)/\Delta)^2}}{\overline{J(\Delta)/\Delta}} } + O(\beta^2\overline{\Delta^2}).
\end{align}
Closed expressions for $h_\infty$ can only be obtained for certain bath models. Usually in the context of the RWA the imaginary part $h_t$ of the bath correlation function is neglected at all as it only modifies the coherent dynamics by shifting the eigenenergies as seen in \cref{eq:RWA} in the main text. However, the contribution to the steady-state beyond the zero coupling limit depends on $h_\infty$  \cite{JThingaPHaenggi2012} and thus it cannot be neglected. \newline
Let us now focus on the special case of an Ohmic spectral density \cref{eq:ohmic} and take the limit $E_c\to\infty$ wherever it is possible. In this way we get universal expressions that are valid independent of how the cutoff is introduced. \newline
First of all the damping kernel $h_\infty(0)$ in \cref{eq:dampvacth} does neither contribute to the variance of $h_\infty$ nor to the weight of the pseudo-Lindblad dissipator \cref{eq:pseudolindblad} in the main text. It only provides a coherent contribution in the Lamb-shift Hamiltonian,
\begin{align}
	\hat{H}_\infty^\mathrm{LS} = \hbar\, h_\infty(0) \hat{S}^2 + \dots,
\end{align}
as can be seen from the energy-basis representation of \cref{eq:lambshift,eq:redfielddissipator} in the main text. Secondly, in the limit $E_c\to\infty$ the vacuum fluctuations $h_\infty^\mathrm{vac}(\Delta)$ in \cref{eq:dampvacth} vanish. Thus, only the contribution due to thermal fluctuations in \cref{eq:dampvacth}, $h_\infty^\mathrm{th}(\Delta)$,  enter the weights. Due to the antisymmetry of $h_\infty^\mathrm{th}(\Delta)$ its variance reduces to the average over the squares, i.e.\ $\lim_{E_c\to\infty}V[h_\infty]=\lim_{E_c\to\infty}\overline{(h_\infty^{\mathrm{th}})^ 2}$. Keeping $\beta E_c$ fixed, while taking the limit $E_C\to\infty$, we find
\begin{align}
	&\lim\limits_{\substack{E_c\to\infty \\ \beta\to 0}} h_\infty^\mathrm{th}(\Delta) \overset{\beta E_c=\xi}{=} -\gamma \frac{\Delta}{\hbar}\, \chi, \\
&\chi \equiv \cot(\xi/2)/2 +\xi^2/\pi \sum\limits_{l=1}^\infty \frac{1}{l(\xi^2 - (2\pi l)^2)}.
	\label{eq:thermalfluctuations}
\end{align}
Finally for the Ohmic spectral density $J(\Delta)=\Delta+O(\overline{\Delta^2}/E_c^2)$ the weights reduce to
\begin{align}
	\lVert\hat{A}_\infty^\pm\rVert^2 = \frac{\gamma}{\hbar\beta}&\Bigg\{\pm\Big[ 1 + \beta^2\, \overline{\frac{\Delta^2}{12}}\Big] \notag \\ &+ \sqrt{1 + \beta^2\,\frac{5}{12}\, \overline{\Delta^2} + \beta^2\, \chi^2 \overline{\Delta^2}} \notag \\ &+ O(\beta^4\overline{\Delta^4})\Bigg\} + O(\overline{\Delta^2}/E_c^2),
\end{align}
and thus the relative weight scales with $\beta$ in second order, i.e.\ $\lVert \hat{A}_\infty^-\rVert^2/\lVert \hat{A}_\infty^+\rVert^2 \simeq \beta^2 [1/16+\chi^2/2]  \overline{\Delta^2}$. This is why in the high-temperature limit the negative contribution vanishes and the truncated master equation becomes an exact representation of the Redfield dissipator. 
\section{Importance of an optimized choice of $\lambda_t$ and $\varphi_t$} \label{sec:optimizationDiscussion}
\begin{figure}[b]
	\includegraphics{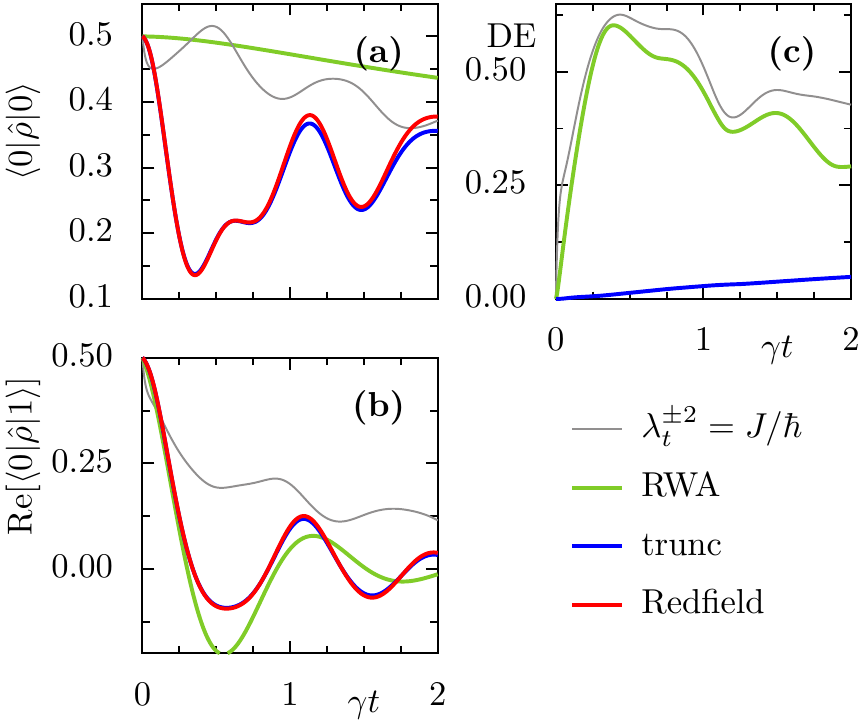}
	\caption{Relevance of the optimization. Dynamics for the different master equations, i.e.\ populations in (a), coherences in (b) and trace distance to the Redfield result in (c). The bath couples to the first site of the chain and the parameters are $l = 5$, $N = 2$, $V = 2J$, $E_c = 17 J$, $\gamma=0.2$ and $T=2J$.}
	\label{fig:dynamics}
\end{figure}
In the main text we motivated the optimal choice of the parameters $\lambda_t$ and $\varphi_t$ that minimize the weight of the negative contribution. In this section we further illustrate the relevance of the optimization procedure. Here we restrict ourselves to a purely real $\lambda_t^+=\lambda_t^-=\lambda_t$ by setting $\varphi_t=0$. By noting that $\sin\varphi_t\propto \tr_\mathrm{S}(\hat{S}\hat{\mathbb{S}})$ and using \cref{eq:trSS}, the optimization is found to reproduce this choice for bath models that do not have a damping kernel nor vacuum fluctuations. In particular this can be achieved by taking into account a bath renormalization Hamiltonian and considering a large cutoff energy. This is discussed for the paradigmatic example of the damped harmonic oscillator in the main text. \newline
Let us now discuss the choice of $\lambda_t$. Note that $\lambda_t^{-2}$ carries the dimension of time, as can be seen from the definition of the newly introduced jump operators $\hat{A}_t^\pm$ in \cref{eq:jumpoperators}. In other words it defines a new timescale and one might ask whether the system's timescale is a reasonable choice. For the extended Hubbard model, which is discussed in the main text, we might choose $\lambda_t^2= J/\hbar$ where $J$ is the tunneling strength between adjacent sites of the chain. In \cref{fig:dynamics} we depict the dynamics of the truncated master equation with optimized $\lambda_t^2$ in blue and for $\lambda_t^{2} = J/\hbar$ in thin grey. The result of the truncated master equation is in very good agreement with the Redfield result in red, since the relative weight of the negative contribution is small. This holds both for the populations in \cref{fig:dynamics} (a) and for the coherence in \cref{fig:dynamics} (b) and also the error measure is particularly small, see \cref{fig:dynamics} (c). In contrast the choice of $\lambda_t^2=J/\hbar$ leads to significant deviations especially for the populations in (a) but also for the coherences in (b). Remarkably, the error is still of the same order as that for the rotating-wave approximation (RWA), see \cref{fig:dynamics} (c). \newline
We can explain our observation further by evaluating the weight of the negative contribution for the choice of $\lambda_t^2 = J/\hbar$, which follows from \cref{eq:weight}. By using the notation of \cref{eq:eigenvalues} of the main text, that is \ $\mathrm{Re} \tr_\mathrm{S}(\hat{S}\hat{\mathbb{S}}_t^\dagger)= \overline{g_t}$ and $\lVert \hat{\mathbb{S}}_t\rVert^2=\overline{g_t^2} + \overline{h_t^2}$, one arrives at
\begin{align}
	\lVert \hat{A}_t^- \rVert ^2  = \Bigg[ \frac{1}{2} - \frac{\overline{g_t}}{ J/\hbar} + \frac{1}{2}\ \frac{\overline{g_t^2} + \overline{h_t^2}}{(J/\hbar)^2} \Bigg] \frac{J}{\hbar},
\end{align}
which for weak coupling reduces to the finite value of $\lVert \hat{A}_t^- \rVert ^2  \simeq J/2\hbar$ independent of the bath parameters. Since $J$ is the typical energy scale of the system this does not correspond to a small value. This explains the bad performance of the new choice. Thus a reasonable choice for the parameter $\lambda_t^2$ is generally not given by the typical timescale of the system. The optimal value of $\lambda_t^2 = \lVert\hat{\mathbb{S}}_t\rVert/\lVert\hat{S}\rVert= (\overline{g_t^2} + \overline{h_t^2})^{1/2}$ is instead determined by the timescale that is related to the amplitude of the bath correlation function.
\section{Multiple baths and nonhermitian coupling} \label{sec:multiplebaths} In the main text we emphasize the relevance of the truncated master equation for the nonequilibrium steady state, when the system is coupled to multiple baths of different temperature. Let us, therefore, consider the case where the total Hamiltonian of the system-bath compound reads 
\begin{align}
	\hat{H}_\mathrm{tot} = \hat{H}_\mathrm{S} + \sum_\alpha (\hat{S}_\alpha \otimes \hat{B}_\alpha + \hat{H}_\mathrm{B,\alpha}),
\end{align}
where $\alpha$ labels different baths. In this section we describe how the truncated master equation, in particular the decomposition of \cref{eq:pseudolindblad,eq:jumpoperators} of the main text, has to be understood in this general scenario. \newline
It is important to note that bath operators for different indices $\alpha \ne \alpha’$ remain uncorrelated because the total Hamiltonian does not contain any cross terms. Consequently the Redfield equation has the very same structure according to  \cref{eq:lambshift,eq:redfielddissipator}, except it involves a sum over the individual coupling operators. For the pseudo-Lindblad equation the decomposition \cref{eq:pseudolindblad,eq:jumpoperators} of the main text simply has to be done for each term independently,
\begin{equation}
	\hat{A}^{\pm}_\alpha(t) = \frac{1}{\sqrt{2 \cos \varphi_\alpha(t) }}  \Big[\lambda_\alpha^\pm(t)\ \hat{S}_\alpha\pm \frac{1}{\lambda_\alpha^\pm(t)} \hat{\mathbb{S}}_\alpha(t)\Big],
\end{equation}
where $\lambda_\alpha^2(t)=\lVert \hat{\mathbb{S}}_\alpha(t) \rVert/\lVert \hat{S}_\alpha \rVert$ and $\sin\varphi_\alpha(t)= \mathrm{Im} \tr_\mathrm{S}(\hat{S}_\alpha\hat{\mathbb{S}}_\alpha^\dagger(t))/\lVert \hat{S}_\alpha \rVert \lVert \hat{\mathbb{S}}_\alpha(t) \rVert$ are given by the optimization procedure. Here we changed the notation and  wrote the time-dependence as argument to not confuse it with the index $\alpha$ that labels the different coupling operators. The optimal parameters minimize the relative weight of the negative contribution for each coupling operator individually. Finally, in the truncated master equation all negative contributions are neglected. \newline
Such a decomposition also holds for non-hermitian coupling 
\begin{align}
	\hat{H}_\mathrm{SB} = (1/2) (\hat{S}\otimes\hat{B} + \hat{S}^\dagger \otimes \hat{B}^\dagger).
\end{align}
Essentially one obtains two channels $\alpha=1,2$ with $\hat{S}_1 = \hat{S}_2^\dagger = \hat{S}$ and the distinct convolution operators, 
\begin{align}
	\hat{\mathbb{S}}_1(t) &= \int\limits_0^t \frac{\tr_\mathrm{B}( \hat{B}^\dagger(\tau) \hat{B} )}{2\hbar^2}\ \hat{S}(-\tau)\ \mathrm{d}\tau, \\
	\hat{\mathbb{S}}_2(t) &= \Bigg[ \int\limits_0^t \frac{\tr_\mathrm{B} (\hat{B}  \hat{B}^\dagger(\tau))}{2 \hbar^2}\ \hat{S}(-\tau)\ \mathrm{d}\tau \Bigg]^\dagger.
\end{align}
For consistency in the hermitian case for $\hat{S}=\hat{S}^\dagger$ and $\hat{B}=\hat{B}^\dagger$ it collapses to one channel with $\hat{\mathbb{S}}_1(t) + \hat{\mathbb{S}}_2(t)= \int_0^{t} C(\tau)\ \hat{S}(-\tau)\ d\tau$ by noting $( \tr_\mathrm{B}( \hat{B}(\tau) \hat{B}) + \tr_\mathrm{B}(\hat{B}  \hat{B}(\tau))^*)/2\hbar=C(t)$.

\bibliography{ref}

\end{document}